\documentclass[a4paper,12pt]{article}
\usepackage{amssymb,amsmath}
\usepackage{epsfig}

\newcommand{\RR}{{\mathbb{R}}}
\newcommand{\CC}{{\mathbb{C}}}
\newcommand{\ZZ}{{\mathbb{Z}}}
\newcommand{\PP}{{\mathbb{P}}}
\newcommand{\ab}{\bar{a}}
\newcommand{\bb}{\bar{b}}
\newcommand{\zb}{\bar{z}}
\newcommand{\oneb}{\bar{1}}
\newcommand{\twob}{\bar{2}}
\newcommand{\half}{\tfrac{1}{2}}
\newcommand{\quar}{\tfrac{1}{4}}

\newcommand{\ee}{{\mathrm{e}}}
\newcommand{\tr}{\mathop{\mathrm{tr}}\nolimits}

\title{Integrable $(2k)$-Dimensional Hitchin Equations}
\author{R.\ S.\ Ward\footnote{email address: richard.ward@durham.ac.uk}
  \bigskip
  \\Department of Mathematical Sciences,
  \\Durham University, Durham DH1 3LE.}
\date{\today}

\begin{document}

\maketitle

\begin{abstract}
\noindent This letter describes a completely-integrable system of Yang-Mills-Higgs
equations which generalizes the Hitchin equations on a Riemann surface
to arbitrary $k$-dimensional complex manifolds. The system arises as
a dimensional reduction of a set of integrable Yang-Mills equations in $4k$
real dimensions. Our integrable system implies other generalizations
such as the Simpson equations and the non-abelian Seiberg-Witten
equations. Some simple solutions in the $k=2$ case are described.
\end{abstract}


\bigskip



\section{Introduction}
This note concerns completely-integrable systems of Yang-Mills-Higgs
equations, and in particular those which may be viewed as higher-dimensional
generalizations of the two-dimensional Hitchin equations (the self-duality
equations on a Riemann surface). Let us begin
by briefly setting out the notation. We denote
local coordinates on $\RR^n$ by $x^\mu$ with $\mu=1,\ldots,n$.
For simplicity we take the gauge group to be SU(2) throughout.
A gauge potential $A_\mu$ takes values in the Lie algebra ${\mathfrak su}(2)$,
so each of $A_1,\dots,A_n$ is an anti-hermitian $2\times2$ matrix.
The curvature (gauge field) is
$F_{\mu\nu}=\partial_\mu A_\nu - \partial_\nu A_\mu + [A_\mu , A_\nu]$.
A Higgs field $\Phi$ takes values in the Lie algebra, or, if complex, in the
complexified Lie algebra ${\mathfrak sl}(2,\CC)$. Its covariant derivative is
$D_\mu=\partial_\mu \Phi + [A_\mu , \Phi]$, and gauge transformations act
by $\Phi\mapsto\Lambda^{-1}\Phi\Lambda$.

The prototype system is the simplest 2-dimensional reduction \cite{L77}
of the 4-dimensional anti-self-dual Yang-Mills equations
\begin{equation} \label{SDYM4}
  F_{12}+F_{34}=0, \quad F_{13}+F_{42}=0, \quad F_{14}+F_{23}=0.
\end{equation}
This reduction can be written as a conformally-invariant system on the
complex plane $\CC$, or more generally on a Riemann surface \cite{H87},
and is effected as follows.
If we take all the fields to depend only on the coordinates $(x^1,x^2)$, and
we define a complex coordinate $z=x^1+ix^2$ and a complex Higgs field
$\Phi=A_3+iA_4$, then (\ref{SDYM4}) reduces to the Hitchin equations
\begin{equation} \label{H2}
  D_{\zb}\Phi=0, \quad F_{z\zb}+\quar[\Phi,\Phi^*]=0.
\end{equation}
Several higher-dimensional generalizations of (\ref{H2}) have been
introduced and studied over the years. But most such generalizations lack
a notable property of the original system (\ref{H2}), namely its complete
integrability. The purpose of this note is to describe some features, and some
solutions, of an integrable $(2k)$-dimensional generalization of (\ref{H2}).

Let us focus specifically on generalizations to $2k$ real (or $k$
complex) dimensions which involve $2k$ real (or $k$ complex) Higgs fields.
Such systems may naturally be viewed as dimensional reductions of
pure-gauge systems in $4k$ dimensions, satisfying linear
relations on curvature such as (\ref{SDYM4}). Of greatest interest
are those that have the eigenvalue form \cite{CDFN83}
\begin{equation} \label{Eig}
  F_{\mu\nu} = \half T_{\mu\nu\alpha\beta} F_{\alpha\beta},
\end{equation}
where $T_{\mu\nu\alpha\beta}$ is totally-skew, because the Bianchi identities
then imply that the gauge field satisfies the second-order Yang-Mills
equations.

Perhaps the best-known example is the \lq octonionic\rq\ system of
\cite{CDFN83}, which has $k=2$. This may be written
\begin{align}
F_{12}+F_{34}+F_{56}+F_{78} &= 0, \nonumber \\
F_{13}+F_{42}+F_{57}+F_{86} &= 0, \nonumber \\
F_{14}+F_{23}+F_{76}+F_{85} &= 0, \nonumber \\
F_{15}+F_{62}+F_{73}+F_{48} &= 0, \nonumber \\
F_{16}+F_{25}+F_{38}+F_{47} &= 0, \nonumber \\
F_{17}+F_{82}+F_{35}+F_{64} &= 0, \nonumber \\
F_{18}+F_{27}+F_{63}+F_{54} &= 0. \label{Octonion}
\end{align}
Whereas the prototype (\ref{SDYM4}) is essentially based on the quaternions,
this system (\ref{Octonion}) is based on the octonions: the components of
$T_{\mu\nu\alpha\beta}$ are constructed from the Cayley numbers.
It is invariant under the group Spin(7), and its
7-dimensional reduction is invariant under $G_2$.
We now reduce to four dimensions by requiring the fields to depend only on
the variables $(x^1,x^2,x^5,x^6)$, defining two complex variables
and two complex Higgs fields by
\begin{equation} \label{CxForm}
z^1=x^1+ix^2, \quad z^2=x^5+ix^6, \quad 
\Phi_1=A_5+iA_6, \quad \Phi_2=A_7+iA_8.
\end{equation}
Then the reduction of (\ref{Octonion}) is
\begin{align}
F_{1\oneb}+F_{2\twob}+\quar[\Phi_1,\Phi_1^*]+
        \quar[\Phi_2,\Phi_2^*] &= 0, \nonumber \\
F_{12}-\quar[\Phi_1,\Phi_2] &= 0,  \nonumber \\
D_{\oneb}\Phi_1-D_2\Phi_2^*=0,  \,\, 
        D_{\twob}\Phi_1+D_1\Phi_2^* &= 0. \label{Oct4}
\end{align}
Here the subscript $1$ in $F_{1\oneb}$ and $D_1$ refers to $z^1$,
whereas $\oneb$ refers to the complex conjugate variable $\zb^1$.
The equations (\ref{Oct4}) are more familiar in the $\RR^4$ (real) form
\begin{equation} \label{KW}
  \left(F-\half[\Phi\wedge\Phi]\right)^+=0, \quad
   \left(D\Phi\right)^-=0, \quad D*\Phi=0,
\end{equation}
where $\Phi=\Phi_\mu dx^\mu$ is a Lie-algebra-valued 1-form formed
from the four real Higgs fields. The `plus' superscript denotes the
self-dual part of a 2-form, and the `minus' superscript the anti-self-dual part.
This system has appeared in several contexts over the years
\cite{DT98, BKS98, KW07, H09, GU12, C15},
and has variously been referred to as the non-abelian Seiberg-Witten equations
or the Kapustin-Witten equations. Known solutions include several obtained
using a generalized 't Hooft ansatz \cite{DH12}.

A different generalization of (\ref{H2}), defined on any K\"ahler manifold,
is one attributed to Simpson \cite{S88}. In $k$ complex dimensions, with
complex coordinates $z^a$, $a=1,\ldots,k$, it takes the form
\begin{align}
F_{1\oneb}+\ldots+F_{k\bar{k}}+\quar[\Phi_1,\Phi_1^*]+\ldots+
      \quar[\Phi_k,\Phi_k^*] &= 0, \nonumber \\
F_{ab}=0, \,\, [\Phi_a,\Phi_b]=0, \,\, D_{\ab}\Phi_b=0. \label{Simp}
\end{align}
Note that for $k=1$, this system reduces to the prototype (\ref{H2}).
For $k=2$, it clearly it implies (\ref{Oct4}).  The converse is
not true in general, but it is if one imposes appropriate global conditions:
in particular for smooth fields on a compact K\"ahler surface, it has recently
been shown that (\ref{Simp}) and (\ref{Oct4}) are equivalent \cite{T15}.


\section{An integrable version}
Another approach to generalizing the basic 4-dimensional system (\ref{SDYM4})
is to look for higher-dimensional versions which are completely-integrable
\cite{W84}. For simplicity, we begin with the case $k=2$. An integrable
8-dimensional Yang-Mills system is
\begin{align}
F_{12}+F_{34} &= F_{56}+F_{78} = 0, \nonumber \\
F_{13}+F_{42} &= F_{57}+F_{86} = 0, \nonumber \\
F_{14}+F_{23} &= F_{76}+F_{85} = 0, \nonumber \\
F_{15} &= F_{26} = F_{37} = F_{48}, \nonumber \\
F_{16} &= F_{52} = F_{83} = F_{47}, \nonumber \\
F_{17} &= F_{28} = F_{53} = F_{64}, \nonumber \\
F_{18} &= F_{72} = F_{36} = F_{54}, \label{A4}
\end{align}
which clearly implies the octonionic equations (\ref{Octonion}).
The system (\ref{A4}) has the symmetry group
$\left[{\rm Sp(1)}\times{\rm Sp(2)}\right]/\ZZ_2\subset{\rm SO(8)}$,
which corresponds to a quaternionic K\"ahler structure \cite{S84}.
The ADHM construction of instantons \cite{ADHM78} generalizes to
this case \cite{S84, CGK85, MS88}. Consider now the reduction
to four dimensions, with the same complex variables (\ref{CxForm})
as before. Then (\ref{A4}) reduces to
\begin{equation} \label{A4red}
D_{\ab}\Phi_b=0, \,\, F_{a\bb}+\quar[\Phi_a,\Phi_b^*]=0, \,\,
[\Phi_a,\Phi_b]=0, \,\, F_{ab}=0, \,\, D_{[a} \Phi_{b]}=0,
\end{equation}
where $a,b\in\{1,2\}$.
This system is even more
overdetermined than~(\ref{Simp}).  So we have a string of implications,
where (\ref{A4red}) implies (\ref{Simp}) implies (\ref{Oct4})
implies the four-dimensional Yang-Mills-Higgs equations (the 
reduction of pure Yang-Mills from eight dimensions).

Generalizing (\ref{A4red}) to $k$ complex dimensions is straightforward:
we simply allow the indices $a,b$ to range from $1$ to $k$.
The system (\ref{A4red}) has a very large symmetry group, since it
involves only the holomorphic structure of the underlying complex manifold.
This becomes clearer if we define
\[
    \Phi=\sum_a \Phi_a\,dz^a
\]
as a $(1,0)$-form with values in the
complexified Lie algebra: then (\ref{A4red}) can be written
\begin{equation} \label{A4red2}
D\Phi=0, \,\, F^{1,1}+\quar[\Phi\wedge\Phi^*]=0, \,\,
[\Phi\wedge\Phi]=0, \,\, F^{2,0}=0,
\end{equation}
where $D$ now denotes the covariant exterior derivative.
By contrast, the less-overdetermined systems (\ref{Simp}) and (\ref{Oct4})
depend on an underlying geometric structure, and have less symmetry.

The system (\ref{A4red}) is completely-integrable by virtue of being the
consistency condition for a \lq Lax $(2k)$-tet\rq, namely
\begin{equation} \label{Laxtet}
\eth_a = D_a+\half\zeta\Phi_a, \quad \eth_{\ab} = D_{\ab}+\half\zeta^{-1}\Phi_a^*,
\end{equation}
where $\zeta$ is a complex parameter. The integrability conditions
\[
  [\eth_a,\eth_b]=0=[\eth_a,\eth_{\bb}]
\]
for all $\zeta$ are equivalent to the equations (\ref{A4red}).


\section{Some solutions}
The aim now is to describe some solutions of (\ref{A4red}); these will
therefore also be solutions of the other systems (\ref{Simp}), and
(\ref{KW}) in the $k=2$ case. The equations~(\ref{A4red}) or (\ref{A4red2})
are defined on any $k$-dimensional complex manifold,
and in general one may also allow singularities. For example, in the
$k=1$ case on a compact Riemann surface of genus $g$,
smooth solutions of (\ref{H2}) exist only when $g\geq2$; on the 2-sphere
and the 2-torus, solutions necessarily have singularities \cite{H87}.
Note that the functions $G_{ab}=\tr(\Phi_a \Phi_b)$ are holomorphic, by
virtue of the equations (\ref{A4red2}). In what follows, we look for solutions
which are smooth on $\CC^2$, and for which $G_{ab}$ is a polynomial
in $z^a$. So they may also be viewed as being defined on the projective
plane $\CC\PP^2$, with a singularity on the line at infinity.

To illustrate, let us first consider the abelian case, with the fields
being diagonal, namely $\Phi_a=\phi_a\sigma_3$, where
$\sigma_3={\rm diag}(1,\,-1)$.
Then the equations (\ref{A4red2}) are easily solved. The gauge field
vanishes, and therefore we may take the gauge potential to vanish as well.
The remaining equations give $\Phi=d\theta$, where $\theta(z^a)$ is an
arbitrary polynomial on $\CC^2$. This is the general abelian solution.

For the non-abelian SU(2) case, we adopt a simplifying ansatz which is
familiar from the lower-dimensional version \cite{HW09}. Namely
let us assume that the gauge potential is diagonal: in other words,
$A_{\ab}=h_{\ab}\sigma_3$. (It should be emphasized that
there are solutions for which this assumption does not hold.)
Then the general local solution is determined by a holomorphic function
$\theta(z^a)$, plus a solution $u=u(\theta,\bar{\theta})$ of the
elliptic sinh-Gordon equation
\begin{equation} \label{SGeqn}
\partial_{\theta} \partial_{\bar{\theta}} \log|u|=\quar\left(|u|^2-|u|^{-2}\right).
\end{equation}
In terms of these, the Higgs fields are given by
\[
  \Phi_a=(\partial_a\theta)
    \left(\begin{matrix} 0 & u \\  u^{-1} & 0\end{matrix}\right),
\]
and the functions determining the gauge potential are
\[
    h_{\ab} = -\half\partial_{\ab}\log(u).
\]
Note that one solution of (\ref{SGeqn}) is $u=1$, but this is effectively
the abelian case of the previous paragraph. In order to get genuine
non-abelian fields, we choose $\theta(z^a)$ to have branch singularities,
and then to get smooth fields one needs $u\neq1$. The simplest such fields
are embeddings of solutions of (\ref{H2}) on $\CC$ into $\CC^2$,
depending on $z^a$ only via a fixed linear combination
$z=\alpha z^1+\beta z^2$. For example, $\theta(z)=z^{3/2}$ gives an
embedding of the \lq one-lump\rq\ solution on $\CC$ \cite{W16}.
Some simple solutions that are not of this embedded type are as follows.

Let $P(z^a)$ be a polynomial of degree at least two, and take
$\theta=\tfrac{2}{3}P^{3/2}$. This gives Higgs fields of the form
\begin{equation} \label{Ansatz}
  \Phi_a=(\partial_aP)
    \left(\begin{matrix} 0 & P\ee^{\psi/2} \\  \ee^{-\psi/2} & 0\end{matrix}\right),
\end{equation}
where $\psi(P,\bar{P})$ satisfies 
\begin{equation} \label{SGeqn2}
\partial_P \partial_{\bar{P}} \psi=\half\left(|P|^2\ee^{\psi}-\ee^{-\psi}\right).
\end{equation}
We now need a smooth solution of (\ref{SGeqn2}) satisfying the boundary
condition $\psi\sim-\log|P|$ as $|P|\to\infty$. There exists a unique such
solution, which is essentially a Painlev\'e-III function \cite{GMN13, W16}.
In fact, if we define $h(t)=t^{-1/3}\ee^{-\psi/2}$, where $t=|P|^{3/2}$,
then (\ref{SGeqn2}) becomes an equation of Painlev\'e-III type, namely
\begin{equation}  \label{PIII}
  h''-\frac{(h')^2}{h}+\frac{h'}{t}+\frac{4}{9h}-\frac{4h^3}{9}=0.
\end{equation}
This has a unique solution with the required asymptotics.

The upshot is that any polynomial $P(z^a)$ gives a solution of (\ref{A4red2})
which is smooth on $\CC^2$ and has
\[
   G_{ab}=\tr\left(\Phi_a\Phi_b\right) = 2P(\partial_aP)(\partial_bP).
\]
It appears (see for example the figure below) that the gauge field $F_{\mu\nu}$
is concentrated around the zero-set of $P$. In the general $k$-complex-dimensional
case, one expects the gauge field to be concentrated around a submanifold of
complex codimension~1, and for the field to be approximately abelian elsewhere.

The simplest case has $P$ quadratic, so that $P(z^a)=0$ is a conic.
\begin{figure}[htb]
\begin{center}
\includegraphics[scale=0.4]{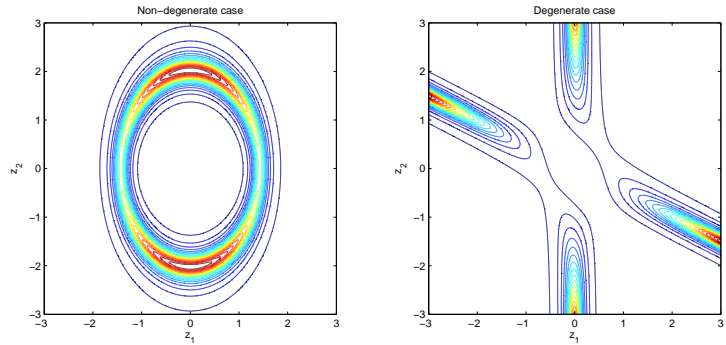}
\caption{Contour plots of the gauge field $|F(z^a)|$ for $z^a\in\RR^2$, with
$P(z^a)=2(z^1)^2+(z^2)^2-4$ and $P(z^a)=z^1(z^1+2z^2)$
respectively.}
\end{center}
\end{figure}
Figure~1 is a plot of the norm $|F|$ of the gauge field, on the real slice
$(z^1,z^2)\in\RR^2$, for the solutions corresponding to the choices
$P(z^a)=2(z^1)^2+(z^2)^2-4$ (on the left), and 
$P(z^a)=z^1(z^1+2z^2)$ (on the right).
Here $|F|$ is computed using the metric
$ds^2 = dz^1\,d\zb^1 + dz^2\,d\zb^2$ on $\CC^2$, which leads
to the formula
\begin{equation}  \label{modF}
  |F|=\left|\ee^{-\psi}-|P|^2\ee^{\psi}\right|
          \left(|\partial_1P|^2+|\partial_2P|^2\right).
\end{equation}
The figures were generated by solving (\ref{PIII}) numerically to get
$\psi$, and then using this formula (\ref{modF}). Clearly $|F|$ is concentrated
around the conic $P(z^a)=0$. The right-hand case corresponds to a degenerate
conic, and is the reduced version of what was called \lq instantons at angles\rq
\,\cite{PT98} for solutions of~(\ref{A4}).

\section{Remarks}
There are some compact complex manifolds $X$ on which smooth
solutions of (\ref{A4red2}) exist. As a trivial example, one could take $X$ to
be a product $S\times X'$, where $S$ is a Riemann surface of genus at least
two, and $X'$ is any other manifold: then a solution of (\ref{H2}) on $S$
is also a solution of (\ref{A4red2}) on $S\times X'$. The moduli space of
solutions on any compact manifold, if it is non-empty, 
has a natural $L^2$ metric, which on general grounds
one expects to be hyperk\"ahler. Even more generally, one could allow
singularities of a specified type, or equivalently for the ambient space to be
non-compact. In this latter case, some of the parameters in the solution space
may have $L^2$ variation, giving rise to a moduli space with a well-defined
metric. Analysing the possible moduli space geometries
which arise in this way would be worthwhile, although a considerable task.

In this note, we have focused on a particular type of reduction of the
integrable system (\ref{A4}), and of its $(4k)$-dimensional generalization.
There are several other dimensional reductions of the octonionic system
(\ref{Octonion}) which are of interest: see, for example, reference \cite{C15}.
In each case, the appropriate reduction of (\ref{A4}) gives an integrable
sub-system, and hence a source of solutions.


\vspace{0.5cm}\noindent{\bf Acknowledgments.}
This work was prompted by a communication from Sergey Cherkis.
The author acknowledges support from the UK Particle Science and Technology
Facilities Council, through the Consolidated Grant ST/L000407/1.


\begin{thebibliography}{99}

\bibitem{ADHM78}
Atiyah, M.\ F., Drinfeld, V.\ G., Hitchin, N.\ J.\ and Manin, Y.\ I.: Construction of
instantons. Phys.\ Lett.\ A {\bf65}, 185--187 (1978).

\bibitem{BKS98}
Baulieu, L., Kanno, H.\ and Singer, I.\ M.: Special quantum field theories
in eight and other dimensions. Commun.\ Math.\ Phys.\ {\bf194}, 149--175 (1998).

\bibitem{C15}
Cherkis, S.\ A.: Octonions, monopoles, and knots.
Lett.\ Math\. Phys.\ {\bf105}, 641--659 (2015).

\bibitem{CDFN83}
Corrigan, E., Devchand, C., Fairlie, D.\ B.\ and Nuyts, J.: First-order
equations for gauge fields in spaces of dimension greater than four.
Nuclear Physics B {\bf214}, 452--464 (1983).

\bibitem{CGK85}
Corrigan, E., Goddard, P.\ and Kent, A.: Some comments on the ADHM construction
in $4k$ dimensions. Commun.\ Math.\ Phys.\ {\bf100}, 1--13 (1985).

\bibitem{DT98}
Donaldson, S.\ K.\ and Thomas, R.\ P.: Gauge theory in higher dimensions. In:
Huggett, S.\ A.\ et al (eds), The Geometric Universe, pp.\ 31--47.
Oxford University Press, Oxford (1998).

\bibitem{DH12}
Dunajski, M.\ and Hoegner, M.: SU(2) solutions to self-duality equations in eight
dimensions. J.\ Geom.\ Phys.\ 62 1747--1759 (2012).

\bibitem{GU12}
Gagliardo, M.\ and Uhlenbeck, K.: Geometric aspects of the Kapustin Witten
equations. J.\ Fixed Point Theory Appl.\ {\bf11}, 185--198 (2012).

\bibitem{GMN13}
Gaiotto, D., Moore, G.\ W.\ and Neitzke, A.: Wall-crossing, Hitchin systems,
and the WKB approximation. Adv.\ Math.\ {\bf234}, 239--403 (2013).

\bibitem{HW09}
Harland, D.\ and Ward, R.\ S.: Dynamics of periodic monopoles.
Physics Letters B {\bf675}, 262--266 (2009).

\bibitem{H09}
Haydys, A.: Gauge theory, calibrated geometry and harmonic spinors.
J.\ London Math.\ Soc.\ {\bf86}, 482--498 (2012).

\bibitem{H87}
Hitchin, N.\ J.: The self-duality equations on a Riemann surface.
   Proc.\ Lond.\ Math.\ Soc.\ {\bf55}, 59--126 (1987).

\bibitem{KW07}
Kapustin, A.\ and Witten, E.: Electromagnetic duality and the geometric Langlands
program. Commun.\ Number Theory Phys.\ {\bf1}, 1--236 (2007).

\bibitem{L77}
Lohe, M.\ A.: Two- and three-dimensional instantons.
Phys.\ Lett.\ B {\bf70}, 325--328 (1977).

\bibitem{MS88}
Mamone Capria, M.\ and Salamon, S.\ M.: Yang-Mills fields on quaternionic spaces.
Nonlinearity {\bf1}, 517--530 (1988).

\bibitem{PT98}
Papadopoulos, G.\ and Teschendorff, A.: Instantons at angles.
Physics Letters B {\bf419}, 115--122 (1998).

\bibitem{S84}
Salamon, S.\ M.: Quaternionic structures and twistor spaces. In:
Willmore, T.\  J.\ and Hitchin, N.\ (eds), Global Riemannian Geometry,
pp.\ 65--74. Ellis Horwood, Chichester (1984).

\bibitem{S88}
Simpson, C.\ T.: Constructing variations of Hodge structure using Yang-Mills
theory and applications to uniformization.
J.\ Amer.\ Math.\ Soc.\ {\bf1}, 867--918 (1988).

\bibitem{T15}
Tanaka, Y.: On the singular sets of solutions to the Kapustin-Witten equations
on compact K\"ahler surfaces. (2015, ArXiv e-prints). arXiv:1510.07739

\bibitem{W84}
Ward, R.\ S.: Completely-solvable gauge-field equations in dimension greater
than four. Nuclear Physics B {\bf236}, 381--396 (1984).

\bibitem{W16}
Ward, R.\ S.: Geometry of solutions of Hitchin equations on $\RR^2$.
Nonlinearity {\bf29}, 756--765 (2016).

\end{thebibliography}
\end{document}